\documentclass{elsart}
\usepackage{natbib}
\usepackage{epsfig}
\begin{document}
\def\crat{C\textsc{iii}]1909/C\textsc{ii}]2326 }
\def\oohb{[O\textsc{ii}]/H$\beta$\,}
\runauthor{K. J. Inskip, P. N. Best \& M. S. Longair}
\begin{frontmatter}
\title{6C radio galaxies at $z \sim 1$: The influence of radio power on 
                  the alignment effect
}
\author[Ap]{Katherine J. Inskip}
\author[roe]{Philip N. Best}
\author[Ap]{Malcolm S. Longair}

\address[Ap]{Astrophysics Group, Cavendish Laboratory}
\address[roe]{IfA, Edinburgh}

\begin{abstract}
Powerful radio galaxies often display enhanced optical/UV
continuum emission and extended emission line regions,
elongated and aligned with the radio jet axis.  The
expansion of the radio source strongly affects the gas
clouds in the surrounding IGM, and the kinematic and
ionization properties of the extended emission line regions
display considerable variation over the lifetime of
individual sources, as well as with cosmic epoch.
We present the results of deep rest-frame UV and optical
imaging and UV spectroscopy of high redshift 6C radio
galaxies.  The interdependence of the host galaxy and
radio source properties are discussed, considering: (i) the
relative contribution of shocks associated with the
expanding radio source to the observed emission line gas
kinematics, and their effect on the ionization state of the
gas; (ii) the similarities and differences between the
morphologies of the host galaxies and aligned emission for a
range of radio source powers; and (iii) the influence of
radio power on the strength of the observed alignment
effect.  

\end{abstract}
\vspace*{-0.4cm}
\end{frontmatter}
\vspace*{-0.4cm}

\section{Introduction}
\vspace*{-0.3cm}

\begin{figure}
\centerline{\vspace*{-0.15cm}
\psfig{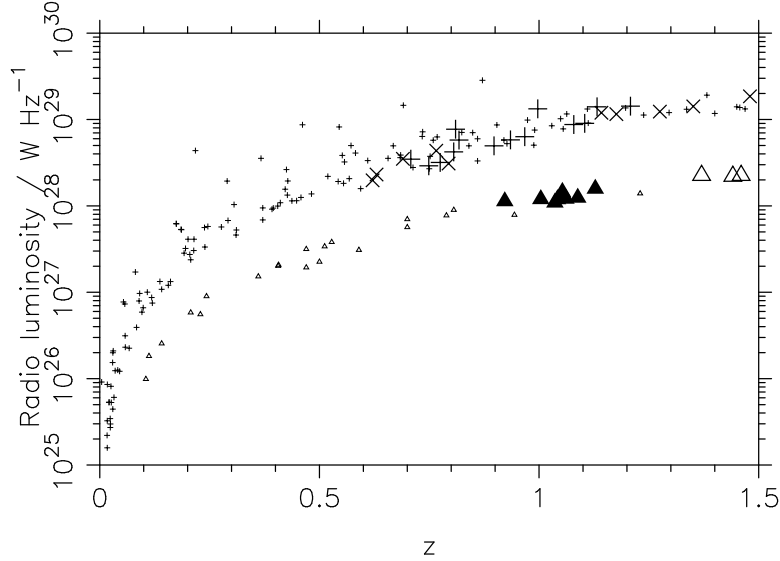} 
}
\caption{\label{pz} $P$--$z$ diagram illustrating the full 3CR and 6CE
samples (small symbols), and the $z \sim 1$ subsamples (large
symbols).  The 6C sources are denoted by triangles, and the
3CR sources  by `$+$' and `$\times$'. (Deep spectra have been obtained
for the galaxies denoted by a filled
triangle or `$+$'.) } 
\end{figure}

At high redshifts, radio galaxies are observed to be surrounded
by extended regions of UV/optical continuum emission, often well aligned
with the radio source axis.
This {\it Alignment Effect} has been well studied in the 3CR sample.
At $z \sim 1$, the aligned structures are typically brighter and more
extensive than at lower redshifts, and more closely aligned
with the radio emission.
A number of different emission mechanisms have been proposed in order
to explain the alignment effect; the most likely models include scattering
of the UV continuum emission from the AGN \cite{UVscat}, jet--induced star
formation \cite{jetSF} and nebular continuum emission
\cite{nebcont}. Additionally, these distant 
radio galaxies also display extended regions of line emission.
One problem with studies of 3CR radio galaxies is that the radio power
of these sources increases with redshift, leading to a
degeneracy between radio power and $z$.
The stronger alignment effect seen in the more powerful higher redshift
objects indicates that the properties of the aligned emission regions
are likely to be closely linked with those of the radio source.  It is
therefore important that the effects of the radio source properties
(size, age and power) on the alignment effect  are fully understood.  
The 6C sample studied in combination with the more powerful 3CR radio
galaxies at the same redshift provides a population of radio galaxies
ideally suited for this investigation, as this enables the $P-z$
degeneracy to be broken.  We have carried out a program of
multiwavelength imaging and spectroscopic observations of a complete
subsample of 11 6C radio sources \cite{6Csam} with flux densities of
2.0\,Jy\,$<\,S_{151}\,<$\,3.93\,Jy, $08^{h}20^{m}<RA<13^{h}01^{m}$,
$34^{\circ}<\delta<40^{\circ}$ and $0.85 < z < 1.5$.
The 6C galaxies are $\sim 6$ times less
powerful radio sources than the 
3CR subsample previously studied by Best et al \cite{blr97b, brl00}; the
sample membership is illustrated in Fig.~\ref{pz}.  

\vspace*{-0.3cm}
\section{The emission line gas}
\vspace*{-0.3cm}

\begin{figure}
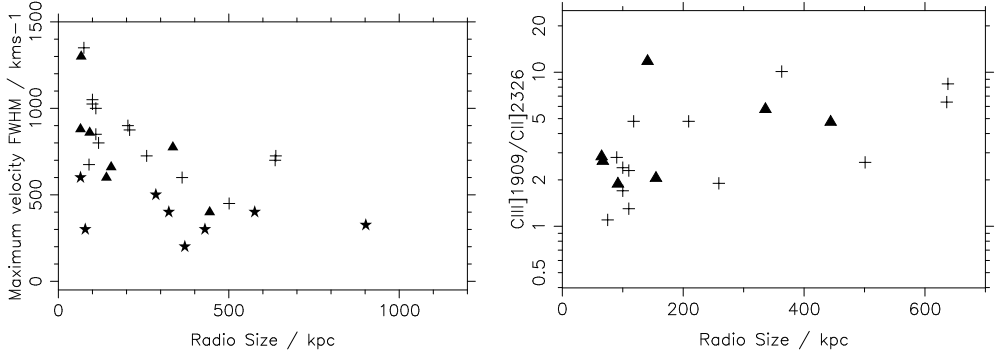

\centerline{
\begin{tabular}{cc}\vspace*{-0.15cm}
\includegraphics[width=11.pc, angle=-90]{InskipFig2a.ps} &
\includegraphics[width=11.pc, angle=-90]{InskipFig2b.ps} 
\end{tabular}}
\caption{\label{spec} (a -- left) The variation of the [O\textsc{ii}]
FWHM with radio size for the $z \sim 1$ 6C and 3CR galaxies (symbols
as in Fig.~\ref{pz}), and a sample of low-$z$ 3CR sources \cite{bm,
inskip02b} (denoted by stars) matched in radio
power to the 6C sources. Partial rank statistics show that FWHM is
independently correlated with radio power, redshift and
 radio size at significance levels of +93\%, +94\%
and $-99$\% respectively. (b -- right) The \crat emission line ratio
vs. radio size for the same sources; these data are correlated with
radio size at a significance level of $>
99$\%.   For both subsamples, the smaller radio galaxies 
exist in a lower ionization state.}
\end{figure}

Deep spectroscopic observations have been carried out for many of the
6C and 3CR radio galaxies in the redshift range $0.85 < z < 1.25$
\cite{brl00,inskip02a}.  Many useful results have been obtained from
this data, including the variation in the ionization state and gas
kinematics (Fig.~\ref{spec}) with radio power, radio size and
redshift.  A statistical analysis of this data shows that the smaller
and more powerful radio sources display more extreme kinematics.  The
spectra of these small sources ($D_{rad} < 120$kpc) are well explained
by   significant levels of shock ionization in addition to
photoionization by the obscured AGN, unlike the larger sources for
which the emission line ratios can be explained simply by AGN
photoionization.  

\vspace*{-0.3cm}
\section{Morphological properties}
\vspace*{-0.3cm}

\begin{figure}
\centerline{\vspace*{-0.25cm}
\psfig{file=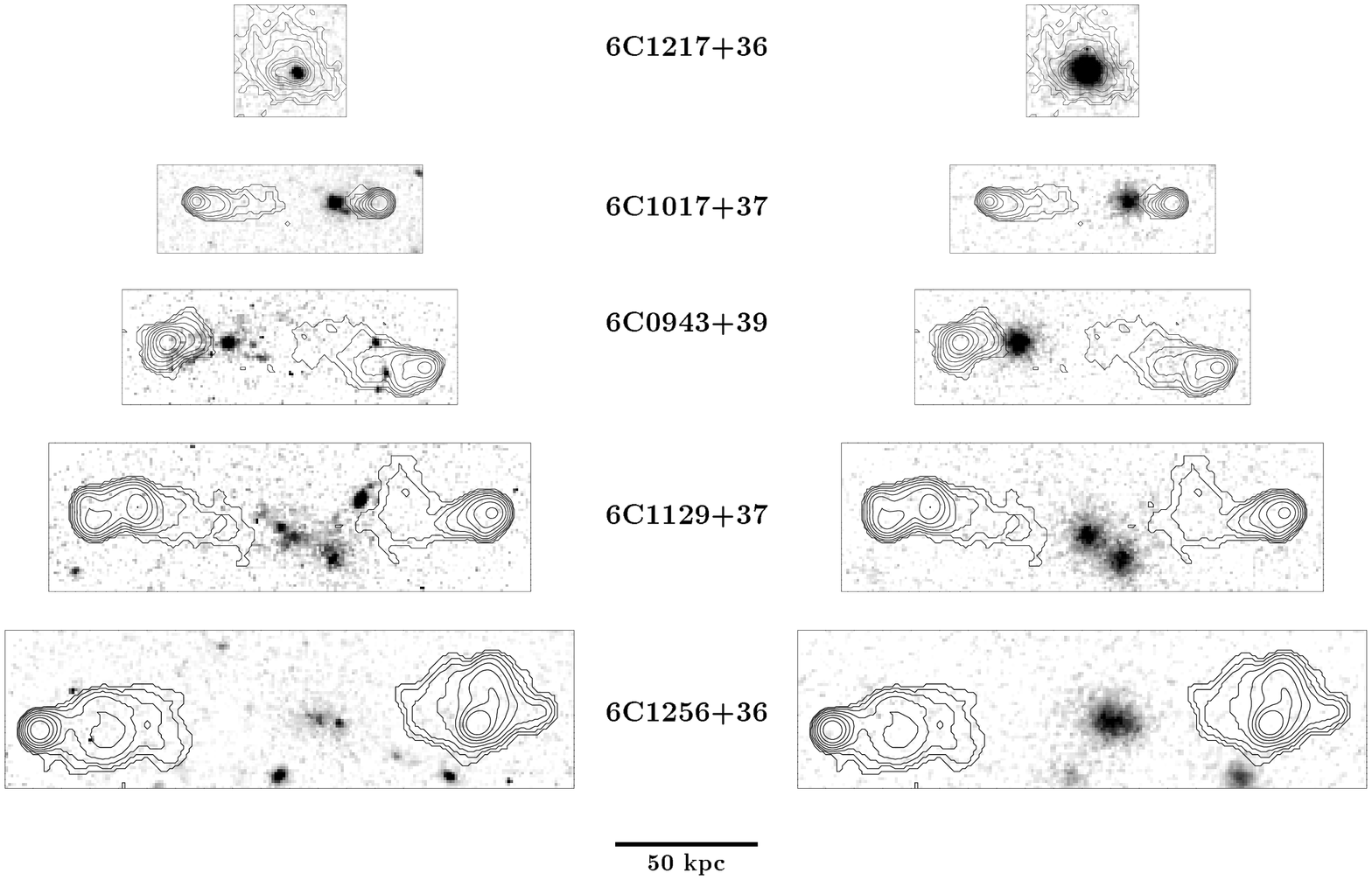,width=9.cm,angle=-90,clip=} 
}
\caption{\label{fig2} HST and UKIRT images of the five smallest of 7
6C radio galaxies in the redshift range $1 < z < 1.3$. Contours represent
the 5GHz VLA observations.}
\end{figure}

\begin{figure}
\centerline{\vspace*{-0.25cm}
\psfig{file=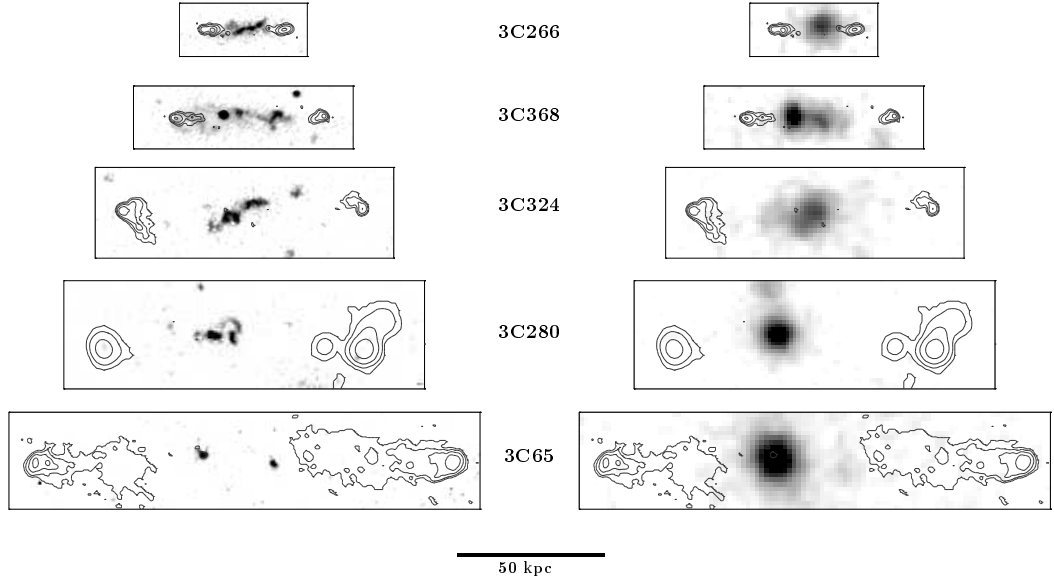,width=7.9cm,angle=-90,clip=} 
}
\caption{\label{fig1} HST and UKIRT images of the five smallest of 8
3CR radio galaxies in the redshift range $1 < z < 1.3$. Contours represent
the 8GHz VLA observations.}
\end{figure}

The imaging observations \cite{thesis} of the 6C galaxies
(Fig.~\ref{fig2}) have 
also produced a number of interesting results.  The UKIRT observations
are dominated by the emission from the old stellar populations and
reveal elliptical host galaxies, whereas the HST observations show a
variety of  morphologies, often with considerable extended
emission aligned with the radio source axis. These structures are
similar to those observed for the 3CR sources \cite{blr97b} at $z \sim 1$
(Fig.~\ref{fig1}), and include arcs and filamentary structures,
diffuse emission and bright knots of emission.   Typically, each
source displays several of these different features, although some
sources (e.g. 6C1019+39, 3C65) appear completely passive, with little 
sign of any extended aligned structures.  Additionally, several
sources in the sample appear to be interacting with close companions. 

Despite the similarities, there are also notable differences in the
aligned structures observed in each sample.  Whilst the smaller 3CR sources
display a clear tendency to have more extensive,
luminous and knotty aligned structures, this is not obvious for the 6C
galaxies at the same redshifts.
The aligned emission surrounding the 6C galaxies is usually also less
extensive.  Additionally, the extended strings of very bright knots of
emission seen around the smaller 3CR radio galaxies are generally not
observed in the 6C sample. 

\vspace*{-0.3cm}
\section{The alignment effect}
\begin{figure}
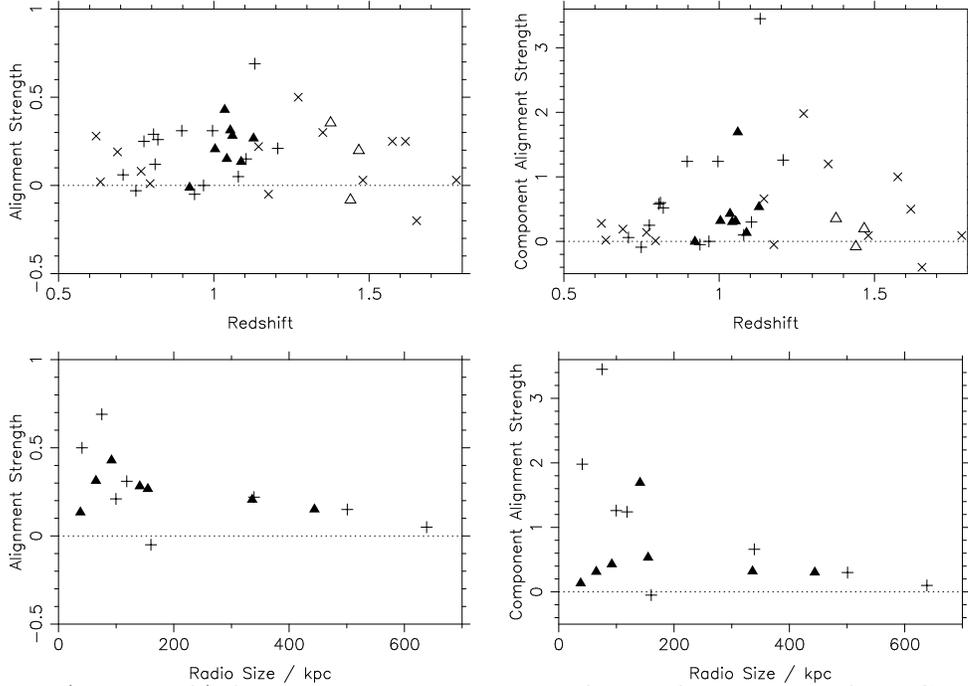

\centerline{\vspace*{-0.3cm}
\begin{tabular}{cc}
\includegraphics[width=10.5pc, angle=-90]{InskipFig5a.ps} &\vspace*{-0.3cm}
\includegraphics[width=10.5pc, angle=-90]{InskipFig5b.ps} \\
\includegraphics[width=10.5pc, angle=-90]{InskipFig5c.ps} &
\includegraphics[width=10.5pc, angle=-90]{InskipFig5d.ps} 
\end{tabular}}
\caption{\label{fig3} (a - top left) Alignment strength $a_s$
vs. $z$ for the full samples of 28 3CR and 11 6C galaxies. (b - top
right) Component alignment strength $a_c$ vs. $z$ for the same data as
in (a).   (c - bottom left) $a_s$ vs. radio size for the 6C and 3CR
radio galaxies in the redshift range $1.0 < z < 1.3$. (d - bottom
right) $a_c$ vs. radio size, for the same data as in (c). Symbols as in Fig.~\ref{pz}.} 
\end{figure}

\vspace*{-0.3cm}
The power of a radio source clearly plays a role in producing the
observed properties of the alignment effect.  In order to investigate
the effects of radio power more fully, we have quantified the
alignment effect in terms of the physical extent of the aligned emission, and the
degree to which it is aligned with the axis of the radio source. 
The {\it Alignment Strength} is given by  $a_s = \epsilon(1 -
\Delta{PA}/45)$, where $\epsilon$ is the ellipticity of the aligned
emission, and $\Delta{PA}$ is the difference in position angle between
the radio and rest-frame UV emission. The {\it ``knottiness''} of the
aligned structures can be considered in terms of the {\it Component
Alignment Strength}, defined as $a_c = N_c a_s$, where $N_c$
represents the number of discrete bright components of emission.
Figure~\ref{fig3} displays the variation in $a_s$ and $a_c$ with
redshift and radio size for the galaxies in both samples.
Both $a_s$ and $a_c$ increase with redshift out to $z=1.1$, reflecting
the increasing importance of the alignment effect at higher redshifts.
A similar trend is also observed in the case of the galaxy colours,
which become bluer with increasing $z$ for both samples.  The decrease
in alignment 
strengths at high redshifts is due to the lower sensitivity of these
observations.  

In order to investigate the effects of radio size on the alignment
strength whilst avoiding confusion with variations with redshift, only
sources within the redshift range $1 < z < 1.3$ are considered.  Overall, both
alignment strength parameters generally decrease with radio size; this
trend is most obvious for the component alignment strength.
Considering the individual samples, $a_s$ and $a_c$ clearly decrease
with radio size for the 3CR sources. However, whilst the variation in
$a_s$ is almost identical for the 6C sources, the component alignment
strengths of the 6C sources are very much weaker, and a clear decrease
in $a_c$ with radio size such as that seen in the case of the more
powerful 3CR radio sources is not observed.  A large value of the
component alignment strength, as seen for the smaller 3CR galaxies,
denotes that the source has a large number of bright knots of
emission, well aligned with the radio source structure.  This result
confirms the observations that such features are present in the
smaller 3CR sources, but not in the less powerful 6C radio sources at
the same redshift.  This is the first obvious effect of the power of
the radio source on the properties of the aligned emission.  

\vspace*{-0.3cm}
\section{Conclusions}
\vspace*{-0.3cm}

From our study of 6C and 3CR sources at $z \sim 1$, we find that although the
size/age of the radio source has the greatest effect on the emission
line gas kinematics and ionization state, radio power is also an
important parameter.  For the
properties of the continuum emission surrounding these sources, radio
size and $z$ are again the most influential parameters. The
alignment effect is more extreme at higher redshifts, and for the
smaller radio sources in the sample; this result is also reflected in
the variation of the galaxy colours.  Radio power does not 
influence alignment strength, and very little difference is seen
between the two $z \sim 1$ samples, except in the case of the smaller 3CR
sources which display more extreme morphologies and noticeably higher
component alignment strengths.   It is clear that different emission mechanisms dominate
under different conditions, suggesting that the importance of
the mechanism which produces the 
bright strings of knots seen in the 3CR sample is directly linked to
radio power.


\begin{thebibliography}{999}

\bibitem{UVscat} Cimatti A., di Serego Alighieri S., Fosbury R.\,A.\,E.,
Salvati M., Taylor D., 1993, MNRAS, 264, 421
\bibitem{jetSF} Chambers K.\,C., Miley G.\,K., van Breugel W.\,J.\,M.,
1987, Nat, 329, 604
\bibitem{nebcont} Osterbrock D.\,E., 1989, {\it Astrophysics of
Gaseous Nebulae and Active Galactic Nuclei}, Mill Valley: University
Science Books 
\bibitem{6Csam} Best P.\,N., Eales S.\,A., Longair M.\,S., Rawlings S.,
R\"{o}ttgering H.\,J.\,A., 1999, MNRAS, 303, 616
\bibitem{blr97b} Best P.\,N., Longair M.\,S., R\"{o}ttgering H.\,J.\,A., 
1997, MNRAS, 292, 758 
\bibitem{brl00} Best P.\,N., R\"{o}ttgering H.\,J.\,A., Longair M.\,S.,
2000, MNRAS, 311, 1 
\bibitem{inskip02a} Inskip K.\,J., Best P.\,N., Rawlings S., Longair M.\,S.,
Cotter G., R\"{o}ttgering H.\,J.\,A., Eales S.\,A., 2002, MNRAS, 337, 1381
\bibitem{bm} Baum S.\,A., McCarthy P.\,J., 2000, ApJ, 119, 2634 
\bibitem{inskip02b} Inskip K.\,J., Best P.\,N., R\"{o}ttgering H.\,J.\,A.,
Rawlings S., Cotter G., Longair M.\,S., 2002, MNRAS, 337, 1407
\bibitem{thesis} Inskip K.\,J., 2002, PhD thesis, Cambridge University


\end{thebibliography}
\end{document}